\documentclass[aps,prd,onecolumn,groupedaddress,showpacs,nofootinbib,amssymb]{revtex4-2}
\usepackage[dvips]{graphicx}
\usepackage{amssymb}
\usepackage{amsmath}
\usepackage{graphicx,,color}
\usepackage{amsfonts}
\usepackage{bm}
\usepackage{cancel}
\usepackage{comment}
\usepackage{floatflt}
\usepackage{slashed}

\newcommand\be{\begin{equation}}
\newcommand\ee{\end{equation}}

\allowdisplaybreaks[4]

\begin{document}

\title{Flat Energy Spectrum of Primordial Gravitational Waves vs Peaks and the NANOGrav 2023 Observation}
\author{V.K. Oikonomou,$^{1,2}$}\email{voikonomou@gapps.auth.gr;v.k.oikonomou1979@gmail.com}
\affiliation{$^{1)}$Department of Physics, Aristotle University of
Thessaloniki, Thessaloniki 54124, Greece \\ $^{2)}$L.N. Gumilyov
Eurasian National University - Astana, 010008, Kazakhstan}

 \tolerance=5000

\begin{abstract}
In this work we present several characteristic examples of
theories of gravity and particle physics scenarios that may yield
an observable energy spectrum of stochastic primordial
gravitational waves, compatible with the 2023 NANOGrav
observations. The resulting theories yield a flat or a peak-like
energy spectrum, and we further seek the conditions which if hold
true, the energy spectrum can be compatible with the recent
NANOGrav stochastic gravitational wave detection. As we show, in
most cases a blue tilted spectrum combined with a relatively low
reheating temperature is needed, the scale of which is determined
by whether the radiation domination era is ordinary or it is an
abnormal radiation domination era. One intriguing Higgs-axion
model, which predicts short slow-roll eras for the axion field at
the post-electroweak breaking epoch, which eventually change the
total equation of state parameter at the reheating era, can
explain the NANOGrav signal, if a blue tilted tensor spectral
index inflationary era precedes the reheating era, and a reheating
temperature of the order $\mathcal{O}(400)\,$GeV. This specific
model produces an energy spectrum of primordial gravitational
waves with a characteristic peak that is detectable from both the
NANOGrav and future LISA experiment, but not from the future
Einstein telescope.
\end{abstract}


\maketitle

\section{Introduction}

The focus in modern theoretical physics experiments is now turned
to the sky and not to terrestrial accelerators. Indeed, the
fundamental theoretical physics concepts, like inflation
\cite{inflation1,inflation2,inflation3,inflation4}, will be
thoroughly tested by gravitational wave experiments
\cite{Hild:2010id,Baker:2019nia,Smith:2019wny,Crowder:2005nr,Smith:2016jqs,Seto:2001qf,Kawamura:2020pcg,Bull:2018lat,LISACosmologyWorkingGroup:2022jok},
and by the stage 4 Cosmic Microwave Background (CMB) experiments
\cite{CMB-S4:2016ple,SimonsObservatory:2019qwx}. The gravitational
wave experiments will directly probe primordial tensor modes which
re-entered the Hubble horizon shortly after the inflationary era,
for huge redshifts and frequencies corresponding to the early
stages of the reheating era (DECIGO, LISA, BBO, Einstein
Telescope).

Recently, the NANOGrav collaboration reported on the first
detection of an isotropic stochastic gravitational wave signal in
a very narrow frequency range in the nHz order, around $f=32\,$nHz
\cite{nanograv} by exploiting the 15 years data set coming from 67
pulsars. The pulsar timing arrays (PTA) are very precise clocks so
the detection of the signal is indeed a striking measurement and
in contrast to the 2020 report of the NANOGrav collaboration, this
2023 report also included the observation of Hellings-Downs
correlation, so the signal is a gravitational wave for sure. The
2023 NANOGrav announcement \cite{nanograv} for a stochastic
gravitational wave background was also confirmed by the EPTA
\cite{Antoniadis:2023ott}, the PPTA \cite{Reardon:2023gzh} and the
CPTA \cite{Xu:2023wog} announcements on the same day. It is not
certain though if the gravitational wave background is of
astrophysical origin coming from supermassive black holes
binaries, or it has a cosmological origin or even if it results
from exotic astrophysical mergers, or even from combinations of
astrophysical and cosmological sources, see the recent review on
prospects of ground detectors with respect to the astrophysical
perspective \cite{Regimbau:2022mdu}. The spectral index of the
NANOGrav observation already significantly constraints the
astrophysical nature of the signal. In this paper we shall
consider the cosmological perspective of the detected signal and
we shall present two classes of models that may explain the
detected signal in the narrow nHz band. The models we shall
present are models of modified gravity
\cite{reviews1,reviews2,reviews3,reviews4,reviews5,Sebastiani:2016ras}
such as $f(R)$ gravity with an abnormal reheating era generated by
geometric terms in the Lagrangian, Einstein-Gauss-Bonnet models
which predict a positive tensor spectral index of the primordial
tensor perturbations, and a model with axion-Higgs higher order
non-renormalizable interactions. Models that belong to the same
class as the $f(R)$ gravity, like single scalar field inflationary
models, predict a flat energy spectrum for the primordial
gravitational waves which can be enhanced only via an abnormal
reheating era generated by geometric terms in the Lagrangian,
while the Einstein-Gauss-Bonnet models predict a positive tensor
spectral index, thus a blue tensor spectrum, and also predict a
peak in the energy spectrum of the primordial gravitational waves.
As we demonstrate the NANOGrav signal can be due to a simple
$f(R)$ gravity inflationary era complemented with a geometrically
generated reheating era, by slow-roll axion disruptions of the
reheating era caused by higher order axion-Higgs interactions,
accompanied with a blue tensor spectrum, or some
Einstein-Gauss-Bonnet theory with significant blue-tilted tensor
spectrum and low-reheating temperature. It is interesting to point
out that the necessity for a blue spectrum is also pointed out in
the recent work \cite{sunnynew}, see also older similar works
\cite{Benetti:2021uea,Vagnozzi:2020gtf}, however in this work we
shall point out that the detection of the signal can also be
explained by higher reheating temperatures, not however in the
context of ordinary Einstein-Gauss-Bonnet theory. The key
ingredient of our approach is that the NANOGrav result can be
explained by the existence of some abnormality during the
reheating era, with relatively low ($\mathcal{O}(400)\,$GeV)
reheating temperature, otherwise, the only possibility for
explaining the NANOGrav signal is described by the recent
\cite{sunnynew}.

\section{Flat Primordial Waves Energy Spectrum: Modified Gravity with an Abnormal Reheating Era}

The motivation for using modifications of general relativity is
multi-fold. Firstly the late-time era can also be described by
modified gravity in its various forms
\cite{reviews1,reviews2,reviews3,reviews4,reviews5}, without
resorting to phantom scalar fields for generating a slightly
phantom dark energy era, and also in the context of modified
gravity, the dark energy era might be dynamical too. Also the
inflationary era can also be described by means of a geometric
theory like $f(R)$ gravity, with the appealing feature of the
geometric description of the inflationary era being the fact that
there is no necessity to account for multiple inflaton couplings
to the Standard Model particles in order to reheat the Universe
post-inflationary. The matter content of a modified gravity like
$f(R)$ gravity is reheated post-inflationary due to curvature
fluctuations after the unstable de Sitter point is reached, with
the most promising model being the $R^2$ model. It is worth
discussing in brief the differences between standard general
relativity action,
\begin{equation}
\label{JGRG6} S_\mathrm{EH}=\int d^4 x \sqrt{-g} \left(
\frac{R}{2\kappa^2} + \mathcal{L}_\mathrm{matter} \right)\, ,
\end{equation}
and the $f(R)$ gravity action,
\begin{equation}
\label{JGRG7} S_{f(R)}= \int d^4 x \sqrt{-g} \left(
\frac{f(R)}{2\kappa^2} + \mathcal{L}_\mathrm{matter} \right)\, .
\end{equation}nd{equation}
where $\mathcal{L}_\mathrm{matter}$ denotes the perfect matter
fluids present. As it can be seen, in the case of $f(R)$ gravity,
the Ricci scalar is replaced by an arbitrary function of the Ricci
scalar. In this case, the field equations can be written in the
standard Einstein-Hilbert way, with the difference that there is
an extra geometric contribution to the energy momentum tensor
caused solely by the terms that contain the function $f(R)$ and
its derivatives, see for example the review \cite{reviews1} for
more details.

In the context of $f(R)$ gravity the energy spectrum of the
primordial gravitational waves is practically flat and well below
the sensitivity curves of all the future interferometers and below
the NANOGrav sensitivity curves. However, there is a possibility
that the $f(R)$ gravity gravitational wave signal is significantly
enhanced, if an abnormal reheating era is generated by non-trivial
powers of the Ricci scalar. From a generic point of view, if some
modification of the standard Einstein-Hilbert gravity controls the
inflationary and late-time dynamics, then it is quite possible
that some intermediate era during the standard reheating and/or
the matter domination era, might shortly be controlled by some
$f(R)$ gravity. Such a possibility was demonstrated in Refs.
\cite{Oikonomou:2022pdf,Odintsov:2022sdk}, so let us here describe
how such a result may be obtained. The general study of primordial
gravitational waves is quite well studied in the literature, see
for example
\cite{Kamionkowski:2015yta,Turner:1993vb,Boyle:2005se,Zhang:2005nw,Caprini:2018mtu,Clarke:2020bil,Smith:2005mm,Giovannini:2008tm,Liu:2015psa,Vagnozzi:2020gtf,Giovannini:2023itq,Giovannini:2022eue,Giovannini:2022vha,Giovannini:2020wrx,Giovannini:2019oii,Giovannini:2019ioo,Giovannini:2014vya,Giovannini:2009kg,Giovannini:2008tm,Kamionkowski:1993fg,Giare:2020vss,Zhao:2006mm,Lasky:2015lej,
Cai:2021uup,Odintsov:2021kup,Benetti:2021uea,Khlopov:2023mpo,Lin:2021vwc,Zhang:2021vak,Visinelli:2017bny,Pritchard:2004qp,Khoze:2022nyt,Casalino:2018tcd,Oikonomou:2022xoq,Casalino:2018wnc,ElBourakadi:2022anr,Sturani:2021ucg,Vagnozzi:2022qmc,Arapoglu:2022vbf,Giare:2022wxq,Oikonomou:2021kql,Gerbino:2016sgw,Breitbach:2018ddu,Schwaller:2015tja,Pi:2019ihn}
and also Ref. \cite{Odintsov:2022cbm} for a recent review on the
topic of extracting the modified gravity effects on primordial
gravitational waves. Let us recall how to extract the overall
effect of $f(R)$ gravity on the general relativistic primordial
gravitational wave waveform, from the present day redshift up to a
redshift $z$, see \cite{Odintsov:2021kup,Odintsov:2022cbm} for
details. The crucial parameter which quantifies the effect of
$f(R)$ gravity on the waveform of general relativity is denoted by
the parameter $a_M$, which for the case of $f(R)$ gravity is,
\begin{equation}\label{amfrgravity}
a_M=\frac{f_{RR}\dot{R}}{f_RH}\, ,
\end{equation}
where $f_{RR}=\frac{d^2f}{d R^2}$ and $f_{R}=\frac{d f}{d R}$, and
the ``dot'' indicates differentiation with respect to the cosmic
time.

The method of extracting and finding the modified gravity effect
and deformation on the general relativistic waveform,  is based on
a WKB approach, which we shall now briefly demonstrate. The
Fourier transformed primordial tensor perturbation $h_{i j}$
satisfies the following evolution equation,
\begin{equation}\label{mainevolutiondiffeqnfrgravity}
\ddot{h}(k)+\left(3+a_M \right)H\dot{h}(k)+\frac{k^2}{a^2}h(k)=0\,
,
\end{equation}
where the parameter $a_M$ is defined in Eq. (\ref{amfrgravity})
for the case of an $f(R)$ gravity. The evolution equation for the
tensor perturbations expressed in terms of the conformal time
$\tau$ takes the following form,
\begin{equation}\label{mainevolutiondiffeqnfrgravityconftime}
h''(k)+\left(2+a_M \right)\mathcal{H} h'(k)+k^2h(k)=0\, ,
\end{equation}
with the ``prime'' indicating differentiation with respect to the
conformal time $\tau$, and also $\mathcal{H}=\frac{a'}{a}$. The
deformation of the primordial gravitational wave waveform
corresponding to the general relativistic case $h_{GR}$ is,
\cite{Nishizawa:2017nef,Arai:2017hxj},
\begin{equation}\label{mainsolutionwkb}
h=e^{-\mathcal{D}}h_{GR}\, ,
\end{equation}
and the parameter $\mathcal{D}$ is,
\begin{equation}\label{dform}
\mathcal{D}=\frac{1}{2}\int^{\tau}a_M\mathcal{H}{\rm
d}\tau_1=\frac{1}{2}\int_0^z\frac{a_M}{1+z'}{\rm d z'}\, .
\end{equation}
Note that the general relativistic waveform $h_{GR}$ appearing in
Eq. (\ref{mainsolutionwkb}) is a solution of the differential
equation (\ref{mainevolutiondiffeqnfrgravityconftime}) in the case
$a_M=0$, and the perturbation of the FRW metric expressed in
conformal time has the following form,
\begin{equation}
  {\rm d}s^{2}=a^{2}[-{\rm d}\tau^{2}+(\delta_{ij}+h_{ij})
  {\rm d}x^{i}{\rm d}x^{j}]\, .
\end{equation}
Hence the modified gravity waveform $h$ describes the deformation
of the general relativistic waveform caused by the presence of
$a_M$. Seeking a WKB solution of Eq.
(\ref{mainevolutiondiffeqnfrgravityconftime}) of the form,
$h_{ij}=\mathcal{A}e^{i\mathcal{B}}h_{ij}^{GR}$, it was shown in
\cite{Nishizawa:2017nef,Arai:2017hxj} that the deformed general
relativistic waveform is,
\begin{equation}\label{mainsolutionwkb}
h=e^{-\mathcal{D}}h_{GR}\, ,
\end{equation}
where $h_{i j}=h e_{i j}$. Thus, the energy spectrum of the $f(R)$
gravity is \cite{Odintsov:2021kup},
\begin{align}
\label{GWspecfR}
    &\Omega_{\rm gw}(f)=e^{-2\mathcal{D}}\times \frac{k^2}{12H_0^2}r\mathcal{P}_{\zeta}(k_{ref})\left(\frac{k}{k_{ref}}
\right)^{n_T} \left ( \frac{\Omega_m}{\Omega_\Lambda} \right )^2
    \left ( \frac{g_*(T_{\rm in})}{g_{*0}} \right )
    \left ( \frac{g_{*s0}}{g_{*s}(T_{\rm in})} \right )^{4/3} \nonumber  \left (\overline{ \frac{3j_1(k\tau_0)}{k\tau_0} } \right )^2
    T_1^2\left ( x_{\rm eq} \right )
    T_2^2\left ( x_R \right )\, ,
\end{align}
where the CMB pivot scale is $k_{ref}=0.002$$\,$Mpc$^{-1}$, while
$n_T$ and $r$ stand for the tensor spectral index of the
primordial tensor perturbations and the tensor-to-scalar ratio.
Hence the total energy spectrum of the $f(R)$ gravity can be
calculated by simply computing the parameter $\mathcal{D}$ for a
given redshift range. As it was shown in previous works
\cite{Oikonomou:2022pdf,Odintsov:2022sdk} for a standard evolution
of the Universe during the reheating and the matter domination era
the amplification factor is trivial and the same applies for the
late-time era even if it is realized by a non-trivial $f(R)$
gravity. The only non-trivial effect on primordial gravitational
waves is obtained if a non-trivial constant equation of state
(EoS) era is geometrically realized by $f(R)$ gravity  during the
reheating era. This is not hard to think in order to be convinced
since, if one accepts that $f(R)$ gravity controls the
inflationary era and the dark energy era, it might also be
possible to control even for a short period of time, some
intermediate era, like a short period during the reheating era. It
was shown that a constant EoS era, with the EoS parameter being
$w$, is realized by the following $f(R)$ gravity,
\begin{equation}
\label{newsolutionsnoneulerssss} F_{w}(R)=\left
[\frac{c_2\rho_1}{\rho_2}-\frac{c_1\rho_1}{\rho_2(\rho_2-\rho_1+1)}\right]R^{\rho_2+1}
+\sum_i
\left[\frac{c_1S_i}{\rho_2(\delta_i+2+\rho_2-\rho_1)}\right]
R^{\delta_i+2+\rho_2}-\sum_iB_ic_2R^{\delta_i+\rho_2}+c_1R^{\rho_1}+c_2R^{\rho_2}\,
,
\end{equation}
where $c_1,c_2$ are integration constants, and $\delta_i$ and
$B_i$ are,
\begin{equation}
\label{paramefgdd}
\delta_i=\frac{3(1+w_i)-23(1+w)}{3(1+w)}-\rho_2+2,\,\,\,B_i=\frac{S_i}{\rho_2\delta_i}
\, ,
\end{equation}
where, $a_1$ and $a_2$, $S_i$ and $A$ are,
\begin{equation}
\label{apara1a2} a_1=\frac{3(1+w)}{4-3(1+w)},\,\,\,
a_2=\frac{2-3(1+w)}{2(4-3(1+w))},\,\,\,S_i=\frac{\kappa^2\rho_{i0}a_0^{-3(1+w_i)}}{[3A(4-3(1+w))]^{\frac{3(1+w_i)}{3(1+w)}}},\,\,\,A=\frac{4}{3(w+1)}
\end{equation}
Hence, from inflation to the dark energy era, there are three
periods controlled by $f(R)$ gravity, as follows,
\[
f(R)=\left\{
\begin{array}{ccc}
  R+\frac{R^2}{6M^2}& R\sim R_I\, ,  \\
  F_w(R) & R\sim R_{PI}\ll R_I\, ,   \\
  F_{DE}(R) & R\sim R_0\ll R_{PI}\, , \\
\end{array}\right.
\]
where $R_I$ is the curvature during inflation ,$R_{PI}$  the
curvature during constant EoS parameter era and also $R_{0}$ is
the curvature at late times. Also for late times, the functional
form of the $f(R)$ gravity $F_{DE}(R)$, must be chosen to be some
viable model, for example \cite{Odintsov:2021kup},
\begin{equation}\label{starobinsky}
F_{DE}(R)=-\gamma \Lambda
\Big{(}\frac{R}{3m_s^2}\Big{)}^{\delta}\, ,
\end{equation}
where $m_s$ in Eq. (\ref{starobinsky}) is
$m_s^2=\frac{\kappa^2\rho_m^{(0)}}{3}$, and also $\rho_m^{(0)}$
denotes the present day energy density of cold dark matter. Also,
the parameter $\delta $ is assumed to be  in the interval
$0<\delta <1$, and also $\gamma$ is some dimensionless parameter,
while $\Lambda$ denotes the present day cosmological constant.
Hence, at early times, the inflationary era is generated by an
$R^2$ gravity, at late times by a viable $f(R)$ gravity model
$F_{DE}(R)$ and also $f(R)$ gravity controls an intermediate era
during reheating. It is hard to find an explicit functional form
of the $f(R)$ gravity which will generate such evolutionary
patches, but one simple example could be,
\begin{align}\label{mainfreffective}
& f(R)=e^{-\frac{\Lambda }{R}}\left(R+\frac{R^2}{6 M^2}
\right)+e^{-\frac{R}{\Lambda }}\tanh
\left(\frac{R-R_{PI}}{\Lambda} \right)F_{DE}(R)
+e^{-\frac{R}{\Lambda }}\tanh \left(\frac{R-R_0}{\Lambda}
\right)\left(F_w(R)-R-\frac{R^2}{6 M^2} \right)\, .
\end{align}
Let us assume that a sudden early dark energy era with constant
EoS parameter $w=-0.35$ is realized purely geometrically by $f(R)$
gravity at the reheating era, and specifically in the redshift
interval $z=[10^{15},4\times 10^{16}]$, so in the temperature
range $T=10^9-10^{11}\,$GeV, deeply in the reheating era. In order
to find the amplification factor for $f(R)$ gravity, we need to
perform the integrals of Eq. (\ref{dform}) for the redshift
intervals $z=[0,z_{pr}]$ and $z=[z_{pr},z_{end}]$, with $z_{pr}=
10^{15}$ and $z_{end}=4\times 10^{16}$. The dominant form of
$f(R)$ is different in these two redshift intervals, thus,
\begin{equation}\label{dformexplicitcalculation}
\mathcal{D}=\frac{1}{2}\left(\int_0^{z_{pr}}\frac{a_{M_1}}{1+z'}{\rm
d z'}+\int_{z_{pr}}^{z_{end}}\frac{a_{M_2}}{1+z'}{\rm d
z'}\right)\, .
\end{equation}
\begin{figure}[h!]
\centering
\includegraphics[width=40pc]{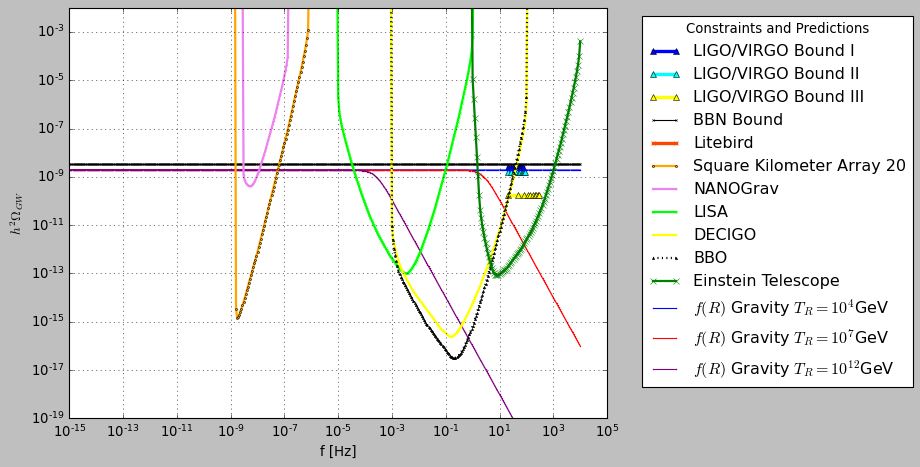}
\caption{The $h^2$-scaled gravitational wave energy spectrum for a
geometrically realized constant EoS $w$-era by $f(R)$ gravity for
the reheating temperature being $T_R=10^{12}$GeV (purple),
$T_R=10^{7}$GeV (red) and to (blue) $T_R=10^{4}$GeV.}
\label{plot1}
\end{figure}
The first integral in Eq. (\ref{dformexplicitcalculation}) is
negligible, but the second integration yields,
$\int_{z_{pr}}^{z_{end}}\frac{a_{M_2}}{1+z'}{\rm d z'}\sim -24$.
Therefore, the overall amplification factor is of the order
$e^{-2\mathcal{D}}\sim \mathcal{O}(10^{10})$, and in Fig.
\ref{plot1} we plot the $h$-scaled energy spectrum of the
primordial gravitational waves for three distinct reheating
temperatures, namely a high $T_R=10^{12}$GeV (purple), an
intermediate reheating temperature $T_R=10^{7}$GeV (red) and a low
reheating temperature $T_R=10^{4}$GeV (blue). Apparently for the
current model, the results presented in Fig. \ref{plot1} indicate
firstly that the predicted gravitational wave signal is compatible
with the NANOGrav observation \cite{nanograv} since it is
detectable and only the purple curve which corresponds to the
highest reheating temperature is excluded because it violates the
LIGO-Virgo constraints. Also we need to note that in the context
of the present model, there is no need to have a significantly low
reheating temperature, which is a mentionable feature indicating
that the reheating temperature requirement for explaining the
NANOGrav results is model dependent. We need now to discuss the
fundamental features of this kind of signals. Firstly it is quite
characteristic since it is quite flat. Thus, this is a
characteristic $f(R)$ gravity spectrum, which is the same as the
single field inflation spectrum, which are flat, but enhanced by a
significant factor, however it still remains quite flat. Also it
is noticeable that the signal can also be detected by the LISA and
the Einstein Telescope in the future. This is the first
characteristic signal which can be obtained by a specific class of
theories, like an $f(R)$ gravity.

\section{Peaks in the Primordial Waves Energy Spectrum I: Einstein-Gauss-Bonnet Theories}

\begin{figure}[h!]
\centering
\includegraphics[width=40pc]{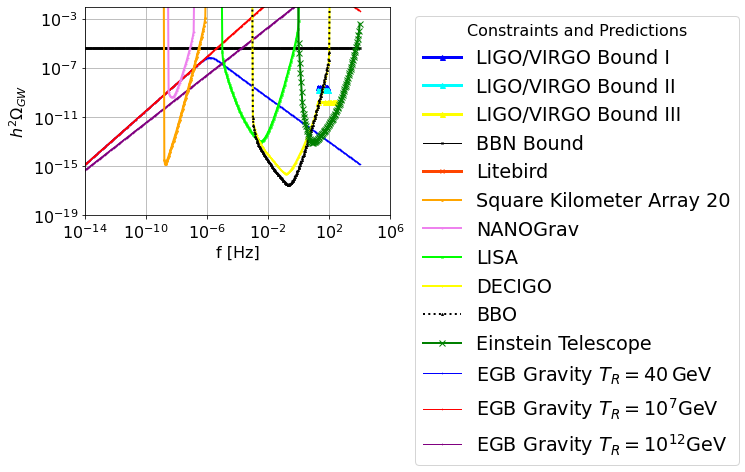}
\caption{The $h^2$-scaled gravitational wave energy spectrum for
Einstein-Gauss-Bonnet theory vs the sensitivity curves of future
gravitational waves experiments for three reheating temperatures,
with a blue tensor spectral index $n_{\mathcal{T}}\sim
0.97$.}\label{plot3}
\end{figure}

Another class of possible observations in the current and future
gravitational waves interferometers is related with peaks in the
energy spectrum of primordial gravitational waves. We shall
commence with pure Einstein-Gauss-Bonnet theories, which have the
characteristic to predict a blue tilted tensor spectral index.
These theories are plagued by the unappealing characteristic of
predicting a non-trivial gravitational wave speed, different from
that of light's in vacuum. This obstacle can be overcome if the
scalar field potential and the Gauss-Bonnet coupling are not free
to chose, but they satisfy a specific relation. Such formalism was
developed in Refs.
\cite{Oikonomou:2022xoq,Oikonomou:2021kql,Odintsov:2020sqy}. The
action for the Einstein-Gauss-Bonnet theory,
\cite{Oikonomou:2022xoq,Oikonomou:2021kql,Odintsov:2020sqy},
\begin{equation}
\label{action} \centering
S=\int{d^4x\sqrt{-g}\left(\frac{R}{2\kappa^2}-\frac{1}{2}\partial_{\mu}\phi\partial^{\mu}\phi-V(\phi)-\frac{1}{2}\xi(\phi)\mathcal{G}\right)}\,
,
\end{equation}
with $R$ denoting the Ricci scalar, $\kappa=\frac{1}{M_p}$ with
$M_p$ being the reduced Planck mass and also $\mathcal{G}$ denotes
the Gauss-Bonnet invariant which is explicitly
$\mathcal{G}=R^2-4R_{\alpha\beta}R^{\alpha\beta}+R_{\alpha\beta\gamma\delta}R^{\alpha\beta\gamma\delta}$
with $R_{\alpha\beta}$ and $R_{\alpha\beta\gamma\delta}$ denoting
the Ricci and Riemann tensors respectively.
\begin{figure}[h!]
\centering
\includegraphics[width=40pc]{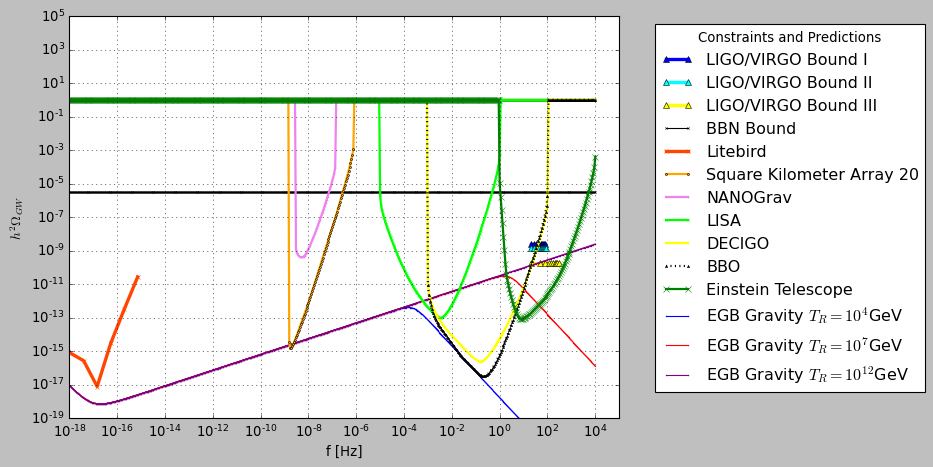}
\caption{The $h^2$-scaled gravitational wave energy spectrum for
Einstein-Gauss-Bonnet theory vs the sensitivity curves of future
gravitational waves experiments for three reheating temperatures,
with a blue tensor spectral index $n_{\mathcal{T}}\sim
0.37$.}\label{plot2}
\end{figure}
Following Refs.
\cite{Oikonomou:2022xoq,Oikonomou:2021kql,Odintsov:2020sqy}, if
the gravitational wave speed is demanded to be equal to unity in
natural units, and thus equal to that of light's, the slow-roll
indices take the following form \cite{Oikonomou:2021kql},
\begin{equation}
\label{index1} \centering \epsilon_1\simeq\frac{\kappa^2
}{2}\left(\frac{\xi'}{\xi''}\right)^2\, ,
\end{equation}
\begin{equation}
\label{index2} \centering
\epsilon_2\simeq1-\epsilon_1-\frac{\xi'\xi'''}{\xi''^2}\, ,
\end{equation}
\begin{equation}
\label{index3} \centering \epsilon_3=0\, ,
\end{equation}
\begin{equation}
\label{index4} \centering
\epsilon_4\simeq\frac{\xi'}{2\xi''}\frac{\mathcal{E}'}{\mathcal{E}}\,
,
\end{equation}
\begin{equation}
\label{index5} \centering
\epsilon_5\simeq-\frac{\epsilon_1}{\lambda}\, ,
\end{equation}
\begin{equation}
\label{index6} \centering \epsilon_6\simeq
\epsilon_5(1-\epsilon_1)\, ,
\end{equation}
with $\mathcal{E}=\mathcal{E}(\phi)$ and $\lambda=\lambda(\phi)$
which are defined in the following way,
\begin{equation}\label{functionE}
\mathcal{E}(\phi)=\frac{1}{\kappa^2}\left(
1+72\frac{\epsilon_1^2}{\lambda^2} \right),\,\, \,
\lambda(\phi)=\frac{3}{4\xi''\kappa^2 V}\, .
\end{equation}
The inflationary observational indices are defined as follows,
\begin{equation}
\label{spectralindex} \centering
n_{\mathcal{S}}=1-4\epsilon_1-2\epsilon_2-2\epsilon_4\, ,
\end{equation}
\begin{equation}\label{tensorspectralindex}
n_{\mathcal{T}}=-2\left( \epsilon_1+\epsilon_6 \right)\, ,
\end{equation}
\begin{equation}\label{tensortoscalar}
r=16\left|\left(\frac{\kappa^2Q_e}{4H}-\epsilon_1\right)\frac{2c_A^3}{2+\kappa^2Q_b}\right|\,
,
\end{equation}
with $c_A$ being the sound speed,
\begin{equation}
\label{sound} \centering c_A^2=1+\frac{Q_aQ_e}{3Q_a^2+
\dot\phi^2(\frac{2}{\kappa^2}+Q_b)}\, ,
\end{equation}
and also with,
\begin{align}\label{qis}
& Q_a=-4 \dot\xi H^2,\,\,\,Q_b=-8 \dot\xi H,\,\,\,
Q_t=F+\frac{Q_b}{2},\\
\notag &  Q_c=0,\,\,\,Q_e=-16 \dot{\xi}\dot{H}\, .
\end{align}
Using the above, the observational indices of inflation can be
significantly simplified in the following way,
\begin{equation}\label{tensortoscalarratiofinal}
r\simeq 16\epsilon_1\, ,
\end{equation}
\begin{equation}\label{tensorspectralindexfinal}
n_{\mathcal{T}}\simeq -2\epsilon_1\left ( 1-\frac{1}{\lambda
}+\frac{\epsilon_1}{\lambda}\right)\, .
\end{equation}
There are various Einstein-Gauss-Bonnet models in the recent
literature \cite{Oikonomou:2021kql} which can be compatible with
the GW170817 event while at the same time provide a viable
phenomenology with a blue tilted tensor spectral index. Without
getting into a detailed analysis of some model here, we shall
consider the model,
\begin{equation}
\label{modelA} \xi(\phi)=\beta  \exp \left(\left(\frac{\phi
}{M}\right)^2\right)\, ,
\end{equation}
where $\beta$ is a dimensionless parameter, and also $M$ is a free
parameter with mass dimensions $[m]^1$. In the formalism of Ref.
\cite{Oikonomou:2021kql}, the scalar potential and the
Gauss-Bonnet coupling are constrained via a non-trivial
differential equation, so for the case at hand the scalar
potential is found to be,
\begin{equation}
\label{potA} \centering V(\phi)=\frac{3}{3 \gamma  \kappa ^4+4
\beta  \kappa ^4 e^{\frac{\phi ^2}{M^2}}} \, ,
\end{equation}
where $\gamma$ is a dimensionless integration constant. It should
be noted that for large field values, decays quite fast, but note
here that the slow-roll conditions are not realized and controlled
solely from the scalar potential, but also for the non-minimal
Gauss-Bonnet coupling function $\xi (\phi)$. For this model, the
slow-roll indices become,
\begin{equation}
\label{index1A} \centering \epsilon_1\simeq \frac{\kappa ^2 M^4
\phi ^2}{2 \left(M^2+2 \phi ^2\right)^2} \, ,
\end{equation}
\begin{equation}
\label{index2A} \centering \epsilon_2\simeq \frac{M^4
\left(2-\kappa ^2 \phi ^2\right)-4 M^2 \phi ^2}{2 \left(M^2+2 \phi
^2\right)^2}\, ,
\end{equation}
\begin{equation}
\label{index3A} \centering \epsilon_3=0\, ,
\end{equation}
\begin{equation}
\label{index5A} \centering \epsilon_5\simeq -\frac{4 \beta  \phi
^2 e^{\frac{\phi ^2}{M^2}}}{\left(M^2+2 \phi ^2\right) \left(3
\gamma +4 \beta  e^{\frac{\phi ^2}{M^2}}\right)} \, ,
\end{equation}
\begin{equation}
\label{index6A} \centering \epsilon_6\simeq -\frac{2 \beta  \phi
^2 e^{\frac{\phi ^2}{M^2}} \left(M^4 \left(2-\kappa ^2 \phi
^2\right)+8 M^2 \phi ^2+8 \phi ^4\right)}{\left(M^2+2 \phi
^2\right)^3 \left(3 \gamma +4 \beta  e^{\frac{\phi
^2}{M^2}}\right)} \, ,
\end{equation}
and the observational indices of inflation are,
\begin{align}\label{spectralpowerlawmodel}
& n_{\mathcal{S}}\simeq -1-\frac{\kappa ^2 M^4 \phi
^2}{\left(M^2+2 \phi ^2\right)^2}+\frac{4 \phi ^2 \left(3 M^2+2
\phi ^2\right)}{\left(M^2+2 \phi ^2\right)^2}\\ & \notag
+\frac{4608 \beta ^2 \phi ^6 e^{\frac{2 \phi ^2}{M^2}} \left(6
\gamma  \phi ^2+16 \beta  e^{\frac{\phi ^2}{M^2}} \left(M^2+\phi
^2\right)+9 \gamma  M^2\right)}{\left(M^2+2 \phi ^2\right)^4
\left(3 \gamma +4 \beta  e^{\frac{\phi ^2}{M^2}}\right)^3} \, ,
\end{align}
regarding the spectral index of scalar perturbations, while
spectral index of tensor perturbations reads,
\begin{align}\label{tensorspectralindexpowerlawmodel}
& n_{\mathcal{T}}\simeq \frac{\phi ^2 \left(-4 \beta e^{\frac{\phi
^2}{M^2}} \left(M^4 \left(3 \kappa ^2 \phi ^2-2\right)+\kappa ^2
M^6-8 M^2 \phi ^2-8 \phi ^4\right)-3 \gamma \kappa ^2 M^4
\left(M^2+2 \phi ^2\right)\right)}{\left(M^2+2 \phi ^2\right)^3
\left(3 \gamma +4 \beta  e^{\frac{\phi ^2}{M^2}}\right)}
 \, .
\end{align}
and finally, the tensor-to-scalar ratio takes the form,
\begin{equation}\label{tensortoscalarfinalmodelpowerlaw}
r\simeq \frac{8 \kappa ^2 M^4 \phi ^2}{\left(M^2+2 \phi
^2\right)^2}\, .
\end{equation}
which yields a tensor spectral index values of the order
$n_{\mathcal{T}}=[0.378856,0.379088]$ and a tensor-to-scalar ratio
$r\sim 0.003$, for $\mu=[22.09147657871,22.09147657877]$,
$\beta=-1.5$, $\gamma=2$, for $N=60$ $e$-foldings. Also we assume
three reheating temperatures, like in the $f(R)$ gravity case,
namely a high reheating temperature $T_R=10^{12}$GeV, an
intermediate reheating $T_R=10^{7}$GeV and a low reheating
temperature $T_R=10^{4}$GeV. In Fig. \ref{plot2} we plot the
$h^2-$scaled energy spectrum of the primordial gravitational waves
for the above Einstein-Gauss-Bonnet model with
$n_{\mathcal{T}}\sim 0.37$ for the three distinct reheating
temperatures. As it can be seen, the spectrum has a characteristic
peak-like form, with the peak being affected by the reheating
temperature and by the tensor spectral index. Also in Fig.
\ref{plot3} we plot the $h^2$-scaled energy spectrum of the
primordial gravitational waves for $n_{\mathcal{T}}\sim 0.97$
(purple curve) and $n_{\mathcal{T}}=1.1$ (blue and red curves).
From both Figs. \ref{plot2} and \ref{plot3} we can see that a
single Einstein-Gauss-Bonnet gravity yields a peak-like signal and
in order for the signal to be detected by the NANOGrav experiment,
it is required that the tensor spectral index must be quite large
of the order $n_{\mathcal{T}}=1.1$ and the reheating temperature
must be quite low of the order $\mathcal{O}(40)\, $GeV, see the
blue curve in Fig. \ref{plot3}. For higher reheating temperatures
and for $n_{\mathcal{T}}=1.1$, the signal violates the BBN bound,
see the red curve in Fig. \ref{plot3}.

In conclusion, Einstein-Gauss-Bonnet theories yield a peak-like
gravitational wave energy spectrum, which can be compatible with
the NANOGrav result if the tensor spectral index is of the order
unity and the reheating temperature is of the GeV order. This
result is compatible with the findings of Ref. \cite{sunnynew}
which also requires a low-reheating temperature and a large blue
tilted tensor spectrum. However, a low-reheating temperature puts
in peril theories which predict primordial gravitational waves via
the electroweak phase transitions. Specifically the first order
electroweak phase transition requires a reheating temperature of
the order of $100\,$GeV, so if the NANOGrav result is correct, a
low reheating temperature can put the electroweak phase transition
theory into peril. This reheating temperature issue however proves
to be model dependent as we show in the next section.

\section{Peaks in the Primordial Waves Energy Spectrum II: Axion-Higgs Higher Order Interactions with a Blue Tilted Tensor Spectral Index}

Another class of models that may yield a peak-like energy spectrum
of primordial gravitational waves involves higher order
non-renormalizable couplings between the axion and the Higgs
particle \cite{Oikonomou:2023bah}. The axion is one among several
currently considered dark matter candidates, for a mainstream of
research and review articles on the axion see
\cite{Preskill:1982cy,Abbott:1982af,Dine:1982ah,Marsh:2015xka,Sikivie:2006ni,Co:2019jts,Co:2020dya,Chen:2022nbb,Oikonomou:2022ela,Roy:2021uye,Tsai:2021irw,Visinelli:2018utg,Oikonomou:2022tux,Odintsov:2020iui,Oikonomou:2020qah,Vagnozzi:2022moj,Banerjee:2021oeu,Machado:2019xuc,Machado:2018nqk}.
It is also mentionable that in the future, experiments will probe
extremely small axion masses in the range $10^{-8}-10^{-16}\,$eV
\cite{Heinze:2023nfb}. In general, couplings between the Higgs and
the axion are quite frequently used in the literature
\cite{Espinosa:2015eda,Im:2019iwd,Dev:2019njv}, but recently
higher order non-renormalizable couplings were considered
\cite{Oikonomou:2023bah}. These higher order non-renormalizable
terms belong to an effective theory that is active at a scale $M$
which is higher than the electroweak scale. Once the electroweak
breaking occurs during the reheating era, when the temperature was
of the order $T\sim 100\,$GeV, the effective operators directly
modify the axion potential, causing new interesting physics in the
axion sector which may have a direct effect on the primordial
gravitational waves energy spectrum, see \cite{Oikonomou:2023bah}
for details. In this paper we shall present in brief the mechanism
described in Ref. \cite{Oikonomou:2023bah}, we shall show that the
energy spectrum of the primordial gravitational waves has a
peak-like form and we shall investigate the conditions needed in
order for the energy spectrum to be detectable from the NANOGrav
collaboration. The result is quite interesting as we now discuss
in brief. Once the electroweak breaking occurs, the axion
potential acquires a new minimum and the axion oscillations around
the origin are destabilized. The axion is free to move along the
potential to reach the new minimum. This can be done in a quick
way, in which case the total EoS during the reheating era is
disturbed and it takes values larger than $w=1/3$, so it is
similar to a short kination era only in the axion dynamics
\cite{Ford:1986sy,Kamionkowski:1990ni,Grin:2007yg,Visinelli:2009kt,Giovannini:1999qj,Giovannini:1999bh,Giovannini:1998bp}.
However, this roll down to the axion new minimum can occur in a
slow-roll way, thus the total EoS during the reheating era is
smaller than $w=1/3$. It is actually this scenario that may yield
a detectable primordial gravitational wave energy spectrum, if
accompanied by a reheating temperature of the order $T\sim
400\,$GeV. Regardless of the way that the axion approaches its new
minimum, once it reaches it, the new vacuum instantly decays to
the Higgs vacuum which is energetically more favorable in the
Universe. We assume that the axion dynamics is described by the
misalignment axion scenario \cite{Marsh:2015xka,Co:2019jts}, in
which case the primordial Peccei-Quinn $U(1)$ symmetry is broken
during inflation and the axion is the radial component of the
primordial complex scalar field that has the $U(1)$ Peccei-Quinn
symmetry. During the inflationary era, the axion has a large
misaligned value, of the order $\phi_i\sim f_a$, with $f_a$ being
the axion decay constant, which is larger than $f_a>10^{9}\,$GeV.
In the misalignment scenario, the axion potential is primordially,
\begin{equation}\label{axionpotentnialfull}
V_a(\phi )=m_a^2f_a^2\left(1-\cos \left(\frac{\phi}{f_a}\right)
\right)\, ,
\end{equation}
hence during inflation and thereafter we have $\phi/f_a<1$, so the
potential is approximately equal to,
\begin{equation}\label{axionpotential}
V_a(\phi )\simeq \frac{1}{2}m_a^2\phi^2\, .
\end{equation}
Once the axion mass becomes of the same order as the Hubble rate,
the axion reaches the minimum of its potential and commences
oscillations around the origin, and redshifts as cold dark matter.

\begin{figure}
\centering
\includegraphics[width=20pc]{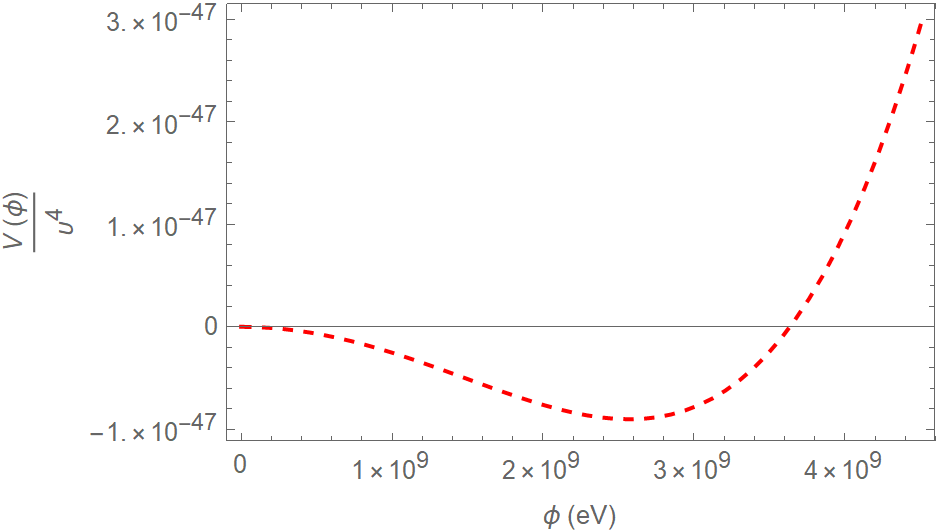}
\includegraphics[width=20pc]{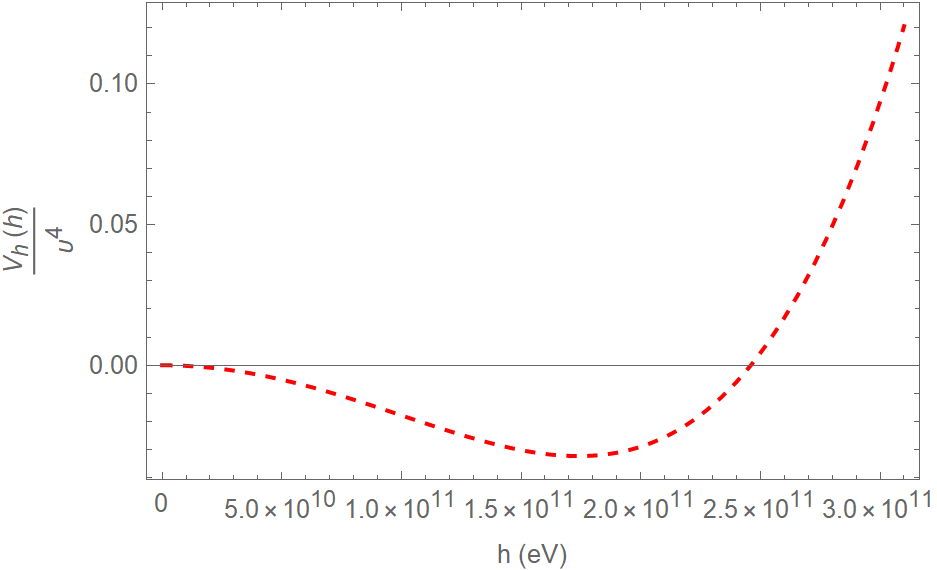}
\caption{The modified axion effective potential for $m_a\sim
10^{-10}$eV, and the effective theory parameters being $M=20\,$TeV
and $\lambda\sim \mathcal{O}(10^{-20})$ and $g\sim
\mathcal{O}(10^{-5})$ (left plot). Also the Higgs potential after
the electroweak symmetry breaking is depicted in the right
plot.}\label{higgsaxion68treeorder}
\end{figure}

\begin{figure}
\centering
\includegraphics[width=19pc]{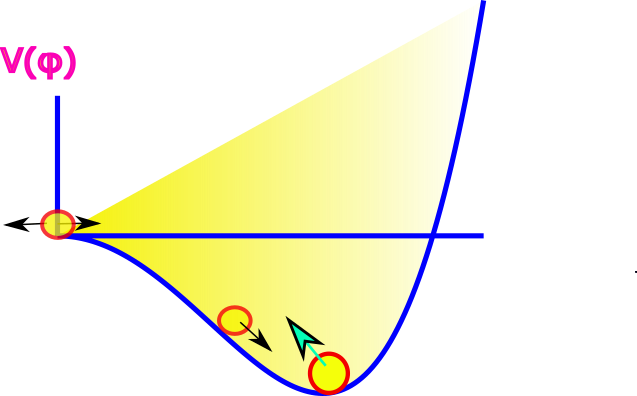}
\caption{The axion scalar potential $V(\phi)$ in the
post-electroweak symmetry breaking epoch develops a deeper minimum
at the direction $(\phi)=(v_s)$. This new minimum disrupts the
axion oscillations at the origin and the axion is free to move
toward to its minimum, either in a slow-roll way or following a
kination trajectory. Once the axion reaches its new minimum, the
new axion scalar minimum in the Universe is not energetically
preferable compared to the Higgs minimum $(h)=(v)$, thus the axion
minimum decays to the Higgs minimum and the potential energy of
the decay is transferred to axion kinetic energy, which makes the
axion to oscillate around the origin in an unbounded way. The
oscillations become destabilized after some time and the same
procedure is repeated perpetually. }\label{pictorialkinationaxion}
\end{figure}

In the present model we shall assume that the axion couples to the
Higgs sector via higher order non-renormalizable operators as
follows,
\begin{equation}\label{axioneightsixpotential}
V(\phi,h)=V_a(\phi)-m_H^2|H|^2+\lambda_H|H|^4-\lambda\frac{|H|^2\phi^4}{M^2}+g\frac{|H|^2\phi^6}{M^4}\,
,
\end{equation}
where $V_a(\phi)$ is defined in Eq. (\ref{axionpotentnialfull}),
and the Higgs field before the electroweak symmetry breaking is
$H=\frac{h+i h_1}{\sqrt{2}}$, with the Higgs mass being $m_H=125$
GeV \cite{ATLAS:2012yve} and $\lambda_H$ stands for the Higgs
self-coupling
$\frac{v}{\sqrt{2}}=\left(\frac{-m_H^2}{\lambda_H}\right)^{\frac{1}{2}}$,
with $v$ being the electroweak symmetry breaking scale
$v\simeq246\,$GeV. The axion mass shall be assumed to be of the
order $m_a\sim 10^{-10}\,$eV and the effective non-renormalizable
operators are assumed to originate from an effective theory with
energy scale of the order $M=20-100\,$TeV\footnote{Notice that in
this section, the parameter $M$ denotes the energy scale of the
effective theory and it is not to be confused with the parameter
$M$ used in the previous section where the Einstein-Gauss-Bonnet
case was considered.} These six and eight dimensional operators
are motivated by the lack of any new particle detection in the LHC
in the energy range $125\,$GeV-15$\,$TeV center of mass. The
Wilson coefficients will be assumed to be $\lambda\sim
\mathcal{O}(10^{-20})$ and $g\sim \mathcal{O}(10^{-5})$. Also, we
shall assume that the electroweak phase transition actually occurs
at $T\sim 100\,$GeV, and it is first order
\cite{Profumo:2007wc,Damgaard:2013kva,Ashoorioon:2009nf,OConnell:2006rsp,Cline:2012hg,Gonderinger:2012rd,Profumo:2010kp,Gonderinger:2009jp,Barger:2008jx,
Cheung:2012nb,Alanne:2014bra,OConnell:2006rsp,Espinosa:2011ax,Espinosa:2007qk,Barger:2007im,Cline:2013gha,Burgess:2000yq,Kakizaki:2015wua,Cline:2012hg,
Enqvist:2014zqa,Barger:2007im,Chala:2018ari,Noble:2007kk,Katz:2014bha}.
After the electroweak phase transition occurs, the Higgs acquires
a vacuum expectation value $H=v+\frac{h+i h_1}{\sqrt{2}}$, and
thus at leading order, the tree order axion potential is modified
and takes the following form,
\begin{equation}\label{axioneffective68}
\mathcal{V}_a(\phi)=V_a(\phi)-\lambda\frac{v^2\phi^4}{M^2}+g\frac{v^2\phi^6}{M^6}\,
,
\end{equation}
where $V_a(\phi)$ is defined in Eq. (\ref{axionpotentnialfull}).
In order to have a concrete idea on the deformation of the axion
potential, after the electroweak symmetry breaking, assume that
$m_a\sim 10^{-10}$eV, $M=20\,$TeV and the Wilson coefficients are
$\lambda\sim \mathcal{O}(10^{-20})$ and $g\sim
\mathcal{O}(10^{-5})$, and we plot the new axion potential in Fig.
\ref{higgsaxion68treeorder} (left plot) and the Higgs effective
potential (right plot). Notice that the Higgs minimum
$(h,\phi)=(v,0)$ is energetically more favorable than the axion
potential minimum $(h,\phi)=(0,v_s)$.
This said behavior for the axion potential after the electroweak
breaking also holds true even if the one-loop corrections are
added in the axionic sector,
\begin{equation}\label{oneloopaxionzerotemperature}
V^{1-loop}(\phi)=\frac{m_{eff}^4(\phi)}{64\pi^2}\left( \ln
\left(\frac{m_{eff}^2(\phi)}{\mu^2}\right)-\frac{3}{2}\right) \, ,
\end{equation}
with $m_{eff}^2(\phi)$,
\begin{equation}\label{axioneffectivemass}
m_{eff}^2(\phi)=\frac{\partial^2 V(\phi,h)}{\partial
\phi^2}=m_a^2-\frac{6 \lambda v^2 \phi^2}{M^2}+\frac{15 g v^2
\phi^4}{M^4}\, ,
\end{equation}
and $\mu$ stands for the renormalization scale. Now since two
different vacua occur in the theory, one the axion new minimum and
the competing Higgs vacuum, the dominant vacuum will be determined
by the depths of the competing vacua, and in our case we have,
\begin{equation}\label{vacuumdecaycondition}
V_h(h)=(v)\gg V(\phi)=(v_s)\, ,
\end{equation}
thus once the axion reaches its minimum, the new vacuum decays
instantly in the Higgs vacuum. A detailed analysis on this can be
found in Ref. \cite{Oikonomou:2023bah}. The instant decay of the
axion scalar field newly developed new vacuum state is unavoidable
since the axion vacuum is competing with the Higgs vacuum state,
thus the Higgs vacuum is energetically more favorable and
therefore the axion vacuum instantly decays to the Higgs vacuum. A
pictorial representation of this procedure can be seen in Fig.
\ref{pictorialkinationaxion}.
Now let us discuss the way that the axion reaches the its new
vacuum. Before the electroweak breaking, the axion behaved as cold
dark matter and it oscillated around the minimum of its potential.
However, in the post-electroweak breaking epoch, the axion
potential develops a new minimum and thus the axion is
destabilized. The oscillations are disrupted and the axion is free
to move along the trajectory in the field space, that leads to the
new minimum. There are two ways for the axion evolution once the
axion oscillations are disrupted, the first is to swiftly evolve
to its new minimum, and the second is to evolve in a slow-roll
way. These two distinct evolutions which have different kinetic
energy for the axion, affect in a different way the total
background EoS of the Universe, which gets disturbed from the
radiation domination value $w=1/3$. In fact, if the axion evolves
swiftly to its new minimum, its kinetic energy dominates over its
potential energy and the axion EoS is described by a stiff EoS
$w_a\simeq 1$, thus the total EoS of the Universe $w$ becomes
larger than the radiation domination value $w=1/3$. This
possibility was examined in Ref. \cite{Oikonomou:2023bah} but by
our analysis performed in this paper, it seems that the NANOGrav
result can be explained only if the axion actually slow-rolls to
its new minimum and the total EoS of the Universe is actually
smaller than the radiation domination value $w=1/3$. So in this
paper we shall study the second scenario, in which the axion
slow-rolls down to its new minimum. Its EoS is thus $w_a\sim -1$,
thus this could perceived as an early dark energy era caused by
the axion and this might significantly affect the total EoS of the
Universe during the radiation domination era. For the purposes of
this paper we shall be more moderate and we shall assume that the
axion slow-roll era affects the total EoS of the Universe in a
minimal way and it changes it from the radiation domination value
$w=1/3$ to $w=0.28$. As we shall see, this plays an important role
for the explanation of the detected gravitational wave signal from
the NANOGrav collaboration.

\begin{figure}[h!]
\centering
\includegraphics[width=40pc]{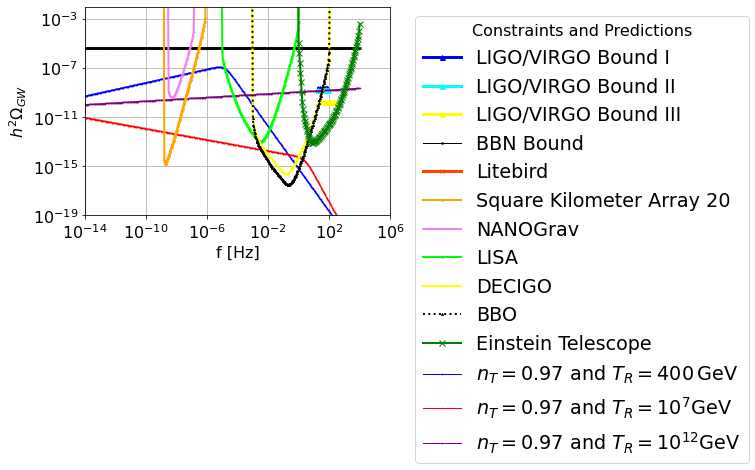}
\caption{The $h^2$-scaled gravitational wave energy spectrum for
and inflationary theory with a blue tensor spectral index, with
three distinct reheating temperatures, for four axion vacuum
decays. The NANOGrav result is verified in the case of a low
reheating temperature of the order $T\sim 400\,$GeV.}\label{plot5}
\end{figure}

The destabilization of the axion during the reheating era can have
observable effects on the energy spectrum of the primordial
gravitational waves via the modes that re-enter the Hubble horizon
at exactly the era at which the axion slow-rolls to its new
minimum. For the purposes of this paper, we shall assume that the
axion slow-roll and decay process occurs two and four times during
the radiation domination era, and at two and four distinct
frequencies. Also we shall assume three different reheating
temperatures, a high reheating temperature $T_R=10^{12}\,$GeV, an
intermediate $T_R=10^7\,$GeV and a low reheating temperature
$T_R=400\,$GeV. The modes that reenter the horizon at a reheating
temperature of the order $T_R=10^7\,$GeV have a wavenumber of the
order $k_R=1.19\times 10^{15}$Mpc$^{-1}$. Therefore, for modes
that reentered the horizon during the era which NANOGrav probes,
should have wavenumbers larger than $k_{t_1}=15\times
10^{5}$Mpc$^{-1}$. So we shall assume that the axion slow-roll and
decay process occurs firstly at two frequencies and specifically
at $k_{t_1}=15\times 10^{5}$Mpc$^{-1}$ , $k_{t_2}=10\times
10^{9}$Mpc$^{-1}$ and secondly at four distinct frequencies at
$k_{t_1}=15\times 10^{5}$Mpc$^{-1}$, $k_{t_2}=6.5\times
10^{9}$Mpc$^{-1}$, $k_{t_3}=10\times 10^{9}$Mpc$^{-1}$ and
$k_{t_4}=15\times 10^{9}$Mpc$^{-1}$. We shall explore the effect
of these decays on the energy spectrum of the primordial
gravitational waves. As we will show, in order for the NANOGrav
result to be explained, one needs a low reheating temperature of
the order $T_R\sim 400\,$GeV, and a blue tilted tensor spectrum
coming from inflation of the order $n_{\mathcal{T}}=0.97$ if the
tensor-to-scalar ratio is of the order $\sim
\mathcal{O}(10^{-3})$. However, if the tensor-to-scalar ratio is
assumed to be larger, the tensor spectral index can be smaller,
and specifically larger than $n_{\mathcal{T}}=0.55$. Such a blue
tilt can come from an Einstein-Gauss-Bonnet theory or some other
inflationary or more exotic scenario. This kind of behavior was
also stressed in the recent work \cite{sunnynew}, but in our case
the required tensor spectral index is smaller and the reheating
temperature can be 10 times larger than the one used in Ref.
\cite{sunnynew}.
\begin{figure}[h!]
\centering
\includegraphics[width=40pc]{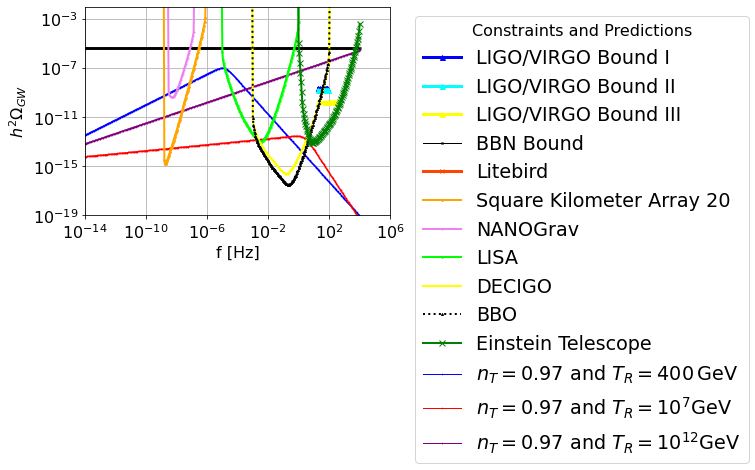}
\caption{The $h^2$-scaled gravitational wave energy spectrum for
and inflationary theory with a blue tensor spectral index, with
three distinct reheating temperatures, for two axion vacuum
decays. The NANOGrav result is verified in the case of a low
reheating temperature of the order $T\sim 400\,$GeV. We assumed a
preceding inflationary era with $n_{\mathcal{T}}=0.97$ and $r\sim
\mathcal{O}(10^{-3})$.}\label{plot4}
\end{figure}
Now, if during the radiation domination era the axion slow-rolls
its potential and the new vacuum decays at some wavenumber
$k_{\mu}$, and if the background EoS is changed from the radiation
domination value $w=1/3$ to some $w$, then the $h^2$-scaled energy
spectrum of the primordial gravitational waves changes by the
following multiplication factor,
 $\sim \left(\frac{k}{k_{\mu}}\right)^{r_c}$, with
$r_c=-2\left(\frac{1-3 w}{1+3 w}\right)$
\cite{Gouttenoire:2021jhk}. For the purposes of this paper we
shall assume that the total EoS parameter is changed from the
$w=1/3$ value to $w=0.28$ two or four distinct times during the
radiation domination era, thus, the $h^2$-scaled energy spectrum
is changed as follows,
\begin{align}
\label{GWspecfRnewaxiondecay}
    &\Omega_{\rm gw}(f)=S_k(f)\times \frac{k^2}{12H_0^2}r\mathcal{P}_{\zeta}(k_{ref})\left(\frac{k}{k_{ref}}
\right)^{n_T} \left ( \frac{\Omega_m}{\Omega_\Lambda} \right )^2
    \left ( \frac{g_*(T_{\rm in})}{g_{*0}} \right )
    \left ( \frac{g_{*s0}}{g_{*s}(T_{\rm in})} \right )^{4/3} \nonumber  \left (\overline{ \frac{3j_1(k\tau_0)}{k\tau_0} } \right )^2
    T_1^2\left ( x_{\rm eq} \right )
    T_2^2\left ( x_R \right )\, ,
\end{align}
with the multiplication factor $S_k(f)$ being,
\begin{equation}\label{multiplicationfactor1}
S_k(f)=\left(\frac{k}{k_{t_1}}\right)^{r_c}\times
\left(\frac{k}{k_{t_2}}\right)^{r_c}\, ,
\end{equation}
for two distinct axion slow-roll and decays, while for four
distinct decays the multiplication factor becomes,
\begin{equation}\label{multiplicationfactor2}
S_k(f)=\left(\frac{k}{k_{t_1}}\right)^{r_c}\times
\left(\frac{k}{k_{t_2}}\right)^{r_c}\times
\left(\frac{k}{k_{t_3}}\right)^{r_c}\times
\left(\frac{k}{k_{t_4}}\right)^{r_c}\, .
\end{equation}
In Figs. \ref{plot4} and \ref{plot5} we have plotted the
$h^2$-scaled energy spectrum of the primordial gravitational waves
with two and four axion decays respectively, for various reheating
temperatures and a blue tilted tensor spectrum with
$n_{\mathcal{T}}=0.97$ and a tensor-to-scalar ratio of the order
$r\sim 0.003$. As it can be seen in both plots, the energy
spectrum of the primordial gravitational waves has a peak-like
form in characteristic frequencies. Regarding the NANOGrav
observation, in order for this axion-Higgs scenario to produce a
detectable gravitational wave signal, it is required that the
tensor-spectral index is blue tilted of the order
$n_{\mathcal{T}}=0.97$ or larger, a low reheating temperature of
the order $T_R\sim 400\,$GeV and two or higher axion slow-roll
evolution and decays during the radiation domination era.
Actually, for the case that four decays take place, the signal is
more pronounced, see for example Fig. \ref{plot5}.
It is mentionable that in both cases in which the signal is
detectable by the NANOGrav collaboration, it is also detectable by
LISA but not from the Einstein Telescope. In conclusion, in the
Higgs-axion model we analyzed in this section, the energy spectrum
of the primordial gravitational waves has a peak-like form at a
specific frequency which depends on when and how many times the
axion slow-rolls its potential and its vacuum decays to the Higgs
vacuum. Moreover, the recent NANOGrav result can be explained if
the inflationary era produces a blue tilted tensor spectral index,
larger than $n_{\mathcal{T}}\geq 0.9$ and if the reheating
temperature is low enough of the order $T_R\sim 400\,$GeV.
Qualitatively similar conclusions regarding the blue tilted tensor
spectrum and the low reheating temperature were reported in Ref.
\cite{sunnynew}, regarding the NANOGrav result. However in our
case, the reheating temperature is low, but not as low as in Ref.
\cite{sunnynew}. In our case, the lowest reheating temperature
that is needed in order to explain the NANOGrav result is four
times larger than the electroweak symmetry breaking temperature.

\section{Conclusions and Discussion}

In this article we presented the various distinct forms of the
energy spectrum of the primordial gravitational waves using
several characteristic theories. We showed that the energy
spectrum can either be flat or it can have a peak-like form. We
used three typical theories that can produce such kind of
spectrums, $f(R)$ gravity, Einstein-Gauss-Bonnet gravity and
finally an inflationary theory with a blue spectrum with abnormal
reheating era caused by axion-Higgs higher order couplings. As we
showed for the case of $f(R)$ gravity, the signal can only be
observed if an abnormal reheating era is generated by geometric
$f(R)$ gravity terms. In this case the spectrum is flat and also
can be detected by all the future gravitational wave experiments,
including the current NANOGrav experiment. Regarding the
Einstein-Gauss-Bonnet theory, only a theory with a blue tilted
tensor spectral index of the order $n_{\mathcal{T}}\sim
\mathcal{O}(1)$ and a low reheating temperature of the order
$T_R\sim \mathcal{O}(40)\,$GeV can be compatible with the NANOGrav
result and also compatible with the BBN bounds. In all the cases,
the Einstein-Gauss-Bonnet class of modified gravity theories
produce a peak-like energy spectrum. Accordingly we examined
another class of theories which leads to an observable energy
spectrum of primordial gravitational waves. This theory contains
higher order non-renormalizable couplings between the Higgs and
the axion and in effect the radiation domination era is disrupted
by short early dark energy axion eras. In order for the signal to
be detected by the NANOGrav experiment, it is important to have a
blue tilted tensor inflationary spectrum and also to have two to
four axion slow-roll eras.

In conclusion, in order for these theories to describe
successfully the NANOGrav signal one must have either one of the
following:
\begin{itemize}
    \item An abnormal reheating era geometrically realized by an
    $f(R)$ gravity.
    \item An abnormal reheating era combined with a relatively low reheating
    temperature $T_R\sim 400\,$GeV and a blue tilted tensor spectral index $n_{\mathcal{T}}\geq 0.97$.
    \item A sufficiently low reheating temperature of the order $T_R\sim
    40\,$GeV and a sufficiently blue tensor spectral index $n_{\mathcal{T}}\geq
    1.1$, regarding Einstein-Gauss-Bonnet theories of gravity.
    This is highly non-trivial to achieve though in the context of
    Einstein-Gauss-Bonnet gravity as it can be shown.
\end{itemize}
Thus in most of the cases, the low reheating temperature, the blue
tilted tensor spectrum and an abnormal reheating era seem to be
the factors that determine whether the model can produce the
NANOGrav signal. The next steps after the astonishing NANOGrav
detection of the first stochastic signal is the commence of the
other two well anticipated experiments, LISA and Einstein
Telescope. Only the joint observations of a stochastic signal may
shed some light on which theory describes the detected stochastic
gravitational wave signal. Also the joint signal can also give
hints on the reheating temperature itself and may also determine
whether the electroweak phase transition ever occurred. Indeed, if
it proven that the reheating temperature never reached 100$\,$GeV,
then this feature would put the theory of electroweak phase
transition in peril, and an alternative mechanism, distinct from a
thermal phase transition, will be needed to explain the
electroweak symmetry breaking.

Finally, it is worth commenting on the cosmological perspective of
the NANOGrav signal. It seems that the spectral slope of the 2023
NANOGrav signal is 3$\sigma$ off the astrophysical prediction
coming from supermassive black holes mergers \cite{sunnynew}. This
feature was also reluctantly discussed in
\cite{Bringmann:2023opz}, based on the 2020 NANOGrav article.
Although specific astrophysical models can be marginally
compatible with the NANOGrav 2023 result \cite{Sampson:2015ada},
for the moment there are too many arguments to be explained for
the astrophysical perspective. Specifically, the complete absence
of anisotropies \cite{NANOGrav:2023gor}, the incomplete
theoretical solution to the final parsec problem
\cite{Sampson:2015ada} which is a theoretical riddle to date, the
absence of single supermassive black hole binaries merger events
in the NANOGrav 2023 announcements, makes us think that from the
Occam's razor perspective, the cosmological perspective of the
2023 signal is more likely, compared to the astrophysical
perspective. This argument is statistically supported currently by
the fact that the many cosmological models are found to provide a
better fit to the NANOGrav 2023 stochastic gravitational wave
background than the astrophysical sources, a better fit which in
Bayes factors ranges from 10 to 100 \cite{NANOGrav:2023hvm}.
However, it is rather too early to make conclusions on the source
of the stochastic gravitational wave background. Surely though, a
decisive conclusion can be made once the spatial anisotropies are
measured, see Ref. \cite{Sato-Polito:2023spo} for a recent study
on this. As expected, this bright new observation has attracted a
lot of attention for new physics explanations
\cite{Cai:2023dls,Han:2023olf,Guo:2023hyp,Yang:2023aak,Addazi:2023jvg,Li:2023bxy,Niu:2023bsr,Yang:2023qlf,Datta:2023vbs,Du:2023qvj,Salvio:2023ynn,Yi:2023mbm,You:2023rmn,Wang:2023div,Figueroa:2023zhu,Choudhury:2023kam,HosseiniMansoori:2023mqh,Ge:2023rce,Bian:2023dnv},
see also
\cite{Schwaller:2015tja,Ratzinger:2020koh,Ashoorioon:2022raz,Choudhury:2023vuj,Choudhury:2023jlt,Choudhury:2023rks,Bian:2022qbh}
for older works which pointed out the ability of NANOGrav to
detect a stochastic gravitational wave signal coming from various
sources. Also it is notable that the stochastic signal detected
may be explained by axion physics too
\cite{Guo:2023hyp,Yang:2023aak}, see also \cite{Machado:2018nqk}
for preceding works.

Lastly, let us further specify the cosmologically viable theories
that can describe the NANOGrav signal. It seems that conventional
theories like Einstein-Gauss-Bonnet gravity requires a large
blue-tilted tensor-spectral index, which is hard to achieve in the
slow-roll approximation. In fact, other theories that also yield
blue-tilted tensor spectral index like string-gas theories
\cite{Kamali:2020drm,Brandenberger:2015kga,Brandenberger:2006pr},
do not yield such a large tensor spectral index, without violating
the Planck 2018 constraints. In fact, the only theory of inflation
which can be potentially compatible with the 2023 NANOGrav signal
is the non-local version of the Starobinsky model
\cite{Calcagni:2020tvw,Koshelev:2020foq,Koshelev:2017tvv}, which
also yields non-Gaussianities of the order unity. Thus although
most theories that yield blue tensor spectral index are incapable
to describe the NANOGrav signal, the non-local version of the
Starobinsky model can provide a significantly large blue tilted,
thus combined with the Higgs-axion model can in principle explain
the NANOGrav 2023 signal. Also the $f(R)$ gravity realized
abnormal reheating era can also explain the NANOGrav signal, but
this is not the end of the story for sure. One cannot be
conclusive at this point for any solution give, phase transition,
cosmic strings or some combinations of inflation with the models
and scenarios we presented in this paper. NANOGrav 2023 was the
start of a new era for physics, the era of the stochastic
gravitational wave background. To fully understand it, it is
compelling to have available the results from the future
experiments like LISA and the Einstein telescope and of course the
Square Kilometer Array, combined with more data from NANOGrav.
Only then we will be sure, if the signal is astrophysical,
cosmological, or a combination of these. Also the shape of the
signal across a large frequency range will only give hints for the
underlying theory. For example a flat signal appearing in all
future experiments, with the same amplitude and energy spectrum
will indicate a theory like the model of $f(R)$ gravity with a
geometrically realized reheating era we described in a previous
section. Or a signal appearing only in the LISA and NANOGrav can
point to another direction. For the moment we are excited to be
blessed to live in the era that the stochastic background is
finally observed and verified. Now a long, exciting and thorny
road awaits till we determine the underlying theory that controls
the cosmos.

\section*{Acknowledgments}

This research has been is funded by the Committee of Science of
the Ministry of Education and Science of the Republic of
Kazakhstan (Grant No. AP19674478).

\end{document}